\documentclass[preprint]{elsarticle}

\usepackage{lineno,hyperref,amssymb}
\modulolinenumbers[5]

\newcommand{\pslash}[1]{#1\!\!\!/\,\,}

\begin{document}

\begin{frontmatter}

  \title{
\vskip-3cm{\baselineskip14pt
\centerline{\normalsize DESY 20--030\hfill ISSN 0418-9833}
\centerline{\normalsize February 2020\hfill}}
\vskip0.5cm
    Bilinear quark operators in the RI/SMOM scheme at three loops}




\author[mymainaddress]{Bernd A. Kniehl}
\ead{kniehl@desy.de}
\address[mymainaddress]{II.~Institut f\"ur Theoretische Physik, Universit\"at Hamburg,\\
    Luruper Chaussee 149, 22761 Hamburg, Germany}
\author[mymainaddress]{Oleg L. Veretin}
\ead{oleg.veretin@desy.de}
\address[mymainaddress]{Institut f\"ur Theoretische Physik, Universit\"at Regensburg,\\
    Universit\"atsstra\ss{}e 31, 93040 Regensburg, Germany}

\begin{abstract}
   We consider the renormalization of the matrix elements of the
    bilinear quark operators $\bar{\psi}\psi$, $\bar{\psi}\gamma_\mu\psi$,
    and $\bar{\psi}\sigma_{\mu\nu}\psi$ at next-to-next-to-next-to-leading order
    in QCD perturbation theory
    at the symmetric subtraction point. This allows us to obtain
    conversion factors between the $\overline{\rm MS}$ scheme and the regularization invariant symmetric momentum subtraction 
    (RI/SMOM) scheme. The obtained results can be used to reduce the errors in
    determinations of quark masses from lattice QCD simulations.
    The results are given in Landau gauge.
\end{abstract}


\end{frontmatter}


\section{Introduction}

The lattice formulation of quantum chromodynamics (QCD)
provides a possibility to estimate long-distance
operator matrix elements from first principles
using Monte Carlo methods. Many important physical observables
can be related to matrix elements of bilinear quark operators
of the form $O_{\mu\dots\nu}=\bar{\psi}\Gamma_{\mu\dots\nu}\psi$, where $\Gamma_{\mu\dots\nu}$
is some Dirac structure that can contain covariant derivatives.

We start from the following expression in Minkowski space:
\begin{equation}
  \int dxdy \,e^{-iq\cdot x-ip\cdot y} \,\langle \psi_{\xi,i}(x) \,O_{\mu\dots\nu}(0)\,
  \bar{\psi}_{\zeta,j}(y) \rangle =
  \delta_{ij} \, S_{\xi\xi'}(-q)
  \Lambda_{\xi'\zeta'}(p,q) S_{\zeta'\zeta}(p)\,,
\label{eq:matrix_element}
\end{equation}  
where $\xi,\zeta$ are spinor indices, $i,j$ are color indices in the fundamental
representation, $S(q)$ is the quark propagator, and $\Lambda(p,q)$ is the amputated Green's
function, which is shown schematically in Fig.~1.
\begin{figure}[h]
\centerline{\includegraphics[width=0.3\textwidth]{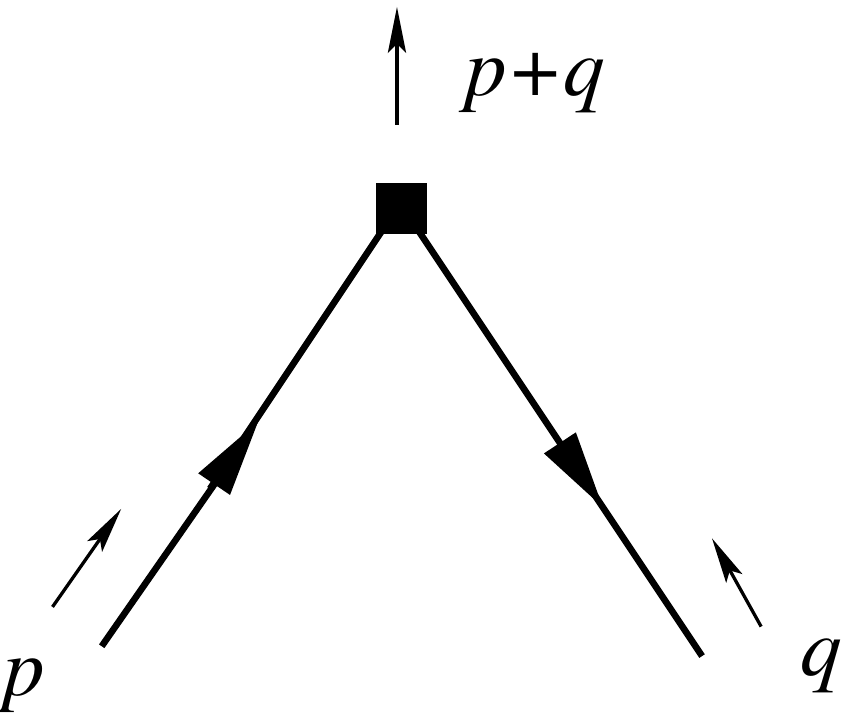}}
\caption{
  Matrix element $\langle \psi(q)\,O(-q-p)\,\bar{\psi}(p) \rangle$
  of a bilinear quark operator in momentum space. The black box denotes the operator, and solid lines denote the external quarks.
  }
\end{figure}

The renormalization of the matrix elements on the lattice is
done in some appropriate renormalization scheme. One of the popular
schemes is the regularization independent momentum subtraction
(RI/MOM) scheme or its variant, the $\rm RI'/MOM$ scheme \cite{Martinelli:1994ty},
where the subtraction is done at the momentum configuration
$p^2=q^2=-\mu^2$, $(p+q)^2=0$.
It has been realized, however, that such a prescription may suffer from a strong sensitivity
to infrared (IR) effects. For example, the pseudoscalar current receives contributions from the
pseoduscalar-meson pole at $(p+q)^2=0$ and is sensitive to condensate
effects of order $O(\Lambda^2_{\rm QCD}/\mu^2)$. 
To avoid such problems, the regularization independent symmetric MOM (RI/SMOM)
scheme has been suggested in Ref.~\cite{Sturm:2009kb}.
The subtraction in this scheme is performed at the symmetric Euclidean point
\begin{equation}
  p^2 = q^2 = (p+q)^2 = -\mu^2 \,, \qquad p\cdot q = \frac{\mu^2}{2} \,.
\label{eq:kinematics}
\end{equation}  
In Eq.~(\ref{eq:kinematics}), none of the four-momenta is exceptional any more, which
provides much better IR behavior for the scheme.

To confront lattice simulations with phenomenological analyses, it is
necessary to convert the matrix elements to the $\overline{\rm MS}$ renormalization
scheme, which is usually adopted in continuum perturbation theory.
The conversion factors for field strengths and masses are know in the RI/MOM scheme through the three-loop order \cite{Chetyrkin:1999pq,Gracey:2003yr}.
The corresponding matching calculations in the RI/SMOM scheme are more involved.
The one-loop results are given in Refs.~\cite{Sturm:2009kb,Gracey:2010ci}. At the two-loop level,
calculations have been done in Refs.~\cite{Almeida:2010ns,Gorbahn:2010bf,Gracey:2011fb} for the
quark currents. The $n=2$ and $n=3$ twist-two operators have been considered
at the two-loop level in Refs.~\cite{Gracey:2011zn,Gracey:2011zg}.
The two-loop singlet axial-vector current has been considered in Ref.~\cite{Gracey:2020rok}.
Recently, in Ref.~\cite{Bednyakov:2020cdf}, the conversion of the strong-coupling constant
has been evaluated at the three-loop order.

The goal of the present work is to evaluate the matching factors between the $\overline{\rm MS}$ and RI/SMOM schemes for the bilinear quark operators in the three-loop approximation.
This paper is organized as follows. In Section~2, we give the underlying definitions and discuss
the main steps of our evaluation procedure. In Section~3, we present our results.
In Section~4, we conclude with a summary.

\section{Evaluation}

In this paper, we consider scalar, vector, and tensor non-singlet bilinear quark operators
in QCD,
\begin{equation}
  J^S = \bar{\psi}\psi\,, \qquad
  J^V_\mu = \bar{\psi}\gamma_\mu\psi\,, \qquad
  J^T_{\mu\nu} = \bar{\psi}\sigma_{\mu\nu}\psi\,,
\end{equation}
where $\sigma_{\mu\nu}=\frac{i}{2}[\gamma_\mu,\gamma_\nu]$, taken at the Euclidean
symmetric kinematic point of Eq.~(\ref{eq:kinematics}).

We adopt the tensor decomposition used in Ref.~\cite{Gracey:2010ci} and write the amputated
Green's functions $\Lambda^{S,V,T}$
in terms of scalar form factors $F^{S,V,T}_j=F^{S,V,T}_j(p,q)$ and the relevant tensor
structures built from the four-momenta $p,\,q$ and Dirac $\gamma$ matrices,
\begin{eqnarray}
\label{eq:LambdaS}
  \Lambda^S(p,q) &=& \Gamma_0 \,\, F^S_1 + \frac{1}{\mu^2}\Gamma_{2,pq} \,\, F^S_2\,, \\
  \Lambda^V_{\mu}(p,q) &=& \gamma_\mu \,\, F^V_1
      + \frac{1}{\mu^2} \pslash{p} p_\mu \,\, F^V_2
      + \frac{1}{\mu^2} \pslash{q} p_\mu \,\, F^V_3
+ \frac{1}{\mu^2} \pslash{p} q_\mu \,\, F^V_4
      \nonumber\\
     &&{}+ \frac{1}{\mu^2} \pslash{q} q_\mu \,\, F^V_5+ \frac{1}{\mu^2} \Gamma_{3,\mu pq} \,\, F^V_6 \,, \\      
  \Lambda^T_{\mu\nu}(p,q) &=& \Gamma_{2,\mu\nu} \,\, F^T_1
      + \frac{1}{\mu^2} (p_\mu q_\nu - p_\nu q_\mu) \,\, F^T_2 
      + \frac{1}{\mu^2} (\Gamma_{2,\mu p} \,p_\nu - \Gamma_{2,\nu p} \,p_\mu) \,\, F^T_3 \nonumber\\
    &&{}+ \frac{1}{\mu^2} (\Gamma_{2,\mu p} \,q_\nu - \Gamma_{2,\nu p} \,q_\mu) \,\, F^T_4
      + \frac{1}{\mu^2} (\Gamma_{2,\mu q} \,p_\nu - \Gamma_{2,\nu q} \,p_\mu) \,\, F^T_5 \nonumber\\
     &&{}+ \frac{1}{\mu^2} (\Gamma_{2,\mu q} \,q_\nu - \Gamma_{2,\nu q} \,q_\mu) \,\, F^T_6
      + \frac{1}{\mu^4} (p_\mu q_\nu - p_\nu q_\mu) \Gamma_{2,pq} \,\, F^T_7
      \nonumber\\
      &&{}+ \frac{1}{\mu^2} \Gamma_{4,\mu\nu pq} \,\, F^T_8 \,,
\label{eq:LambdaT}
\end{eqnarray}
where $\Gamma_n$ denote antisymmetric products of $\gamma$ matrices with the normalization
factor $1/n!$ included, i.e.
\begin{equation}
  \Gamma_0 = \mathbb{I}\,, \qquad 
  \Gamma_{2,\nu_1\nu_2} = \frac{1}{2!}(\gamma_{\nu_1}\gamma_{\nu_2}
           -\gamma_{\nu_2}\gamma_{\nu_1})\,,
  \qquad\mbox{etc.}
\end{equation}  
In Eqs.~(\ref{eq:LambdaS})--(\ref{eq:LambdaT}), we have also used the shorthand notation $p^\alpha \Gamma_{\dots\alpha\dots}=\Gamma_{\dots p\dots}$ for the
contraction of a four-momentum with a tensor.

The evaluation of the above matrix elements is organized as usual, in two
steps: the reduction to master integrals and their evaluation.
After the projection and the evaluation of the color and Dirac traces, we first
reduce the large number of Feynman integrals with the help of
integration-by-parts (IBP) relations \cite{Chetyrkin:1981qh} to a small set of master integrals.
This is done with the help of the computer package FIRE \cite{Smirnov:2014hma}.
Besides the IBP relations, we have additional relations arising
from the symmetric kinematics of Eq.~(\ref{eq:kinematics}).
With these new relations, we can further reduce the number of master integrals.
Finally, we can express all the amplitudes in terms of
2 one-loop, 8 two-loop, and 60 three-loop master integrals.

  Generally, an amplitude can be written as a sum of $N$ master integrals ${\cal M}_j$,
\begin{equation}
  A = \sum_{j=1}^{N} c_j(d) {\cal M}_j \,,
\label{eq:reduced}  
\end{equation}
where the coefficients $c_j(d)$ are rational functions of the space-time dimension $d$ and are
determined in the course of the reduction procedure.

In our case, we have the situation where the master integrals ${\cal M}_j$ have at most $1/\varepsilon^3$
poles in the $\varepsilon$ expansion, while the coefficients $c_j(d)$ can have poles up to $1/\varepsilon^5$.
This means that some of the three-loop master integrals have to be expanded up to $O(\varepsilon^5)$
to extract the finite contributions! If the master integrals are evaluated numerically
with only restricted accuracy (see the discussion below), this is expected to lead to a significant
loss of accuracy in the final renormalized results.

Notice, however, that the choice of the master integrals ${\cal M}_j$ in Eq.~(\ref{eq:reduced}) is
not unique. While the number $N$ of master integrals remains invariant, any
$N$ linearly independent integrals can be chosen as a basis. This freedom can be exploited
to improve the behavior of the coefficients $c_j(d)$ in the limit $d\to4$.
In particular, if all coefficients $c_j(d)$ have no poles as $\varepsilon\to0$, then the set of master integrals
${\cal M}_j$ represents an $\epsilon$-finite basis \cite{Chetyrkin:2006dh,Faisst:2006sr}.
Such a basis can always be constructed as long as no restrictions on the choice of master integrals ${\cal M}_j$ are imposed.
In our evaluation, we choose master integrals with positive indices and work 
with a restricted set of master integrals, namely, the master integrals that are present in the original expressions in Landau gauge
before the reduction procedure. With such conditions, we construct a basis that has at most
$1/\varepsilon$ poles in the coefficients $c_j(d)$.

We now turn to the evaluation of the master integrals. In the kinematics of Eq.~(\ref{eq:kinematics}), we have
single-scale integrals. The integrals through two loops have been considered previously,
in Ref.~\cite{Usyukina:1994iw} using the parametric-integral representation
and, more systematically, in Ref.~\cite{Birthwright:2004kk} using the method of differential equations.
We can briefly summarize our knowledge about the analytic structure of the master integrals in the
kinematics of Eq.~(\ref{eq:kinematics}) as follows. They can be expressed in terms of harmonic
polylogarithms \cite{Remiddi:1999ew} taken at the special point $e^{i\pi/6}$ on the unit
circle in the complex plane. At three loops, we have polylogarithms through weight six.
The basis of relevant constants up to weight six has been constructed in Ref.~\cite{Kniehl:2017ikj}
using differential equations. These master integrals have been considered
recently in Ref.~\cite{Bednyakov:2020cdf}.

In our evaluation, we compute the master integrals numerically instead. For this purpose, we use the method of sector decomposition \cite{Binoth:2000ps,Binoth:2003ak},
which is based on the analytic resolution of singularities and the successive
numerical integration of the parametric integrals by Monte Carlo methods.
For this purpose, we use the implementation of the program package FIESTA \cite{Smirnov:2015mct}.
At the three-loop level, we have four-fold to nine-fold parametric integrals resulting usually
in several hundreds of so-called sector integrals, which are then evaluated numerically
using the program library CUBA \cite{Hahn:2004fe}. With a typical sample of $10^{8}$ function calls,
we achieve a relative accuracy of order $10^{-6}$ for individual master integrals. However, due to
large cancellations between different terms in sums like the one in Eq.~(\ref{eq:reduced}), the resulting
relative accuracy is expected to be worse.

\section{Results and discussion}

In this section, we present our results for the individual form factors defined
in Eqs.~(\ref{eq:LambdaS})--(\ref{eq:LambdaT}).
We perform our evaluations in Landau gauge and for the kinematics in Eq.~(\ref{eq:kinematics}).
After the renormalization of the matrix elements, the $1/\varepsilon^j$ poles cancel.
Since the master integrals are known numerically with restricted accuracy, such
cancellations are only approximate. We demonstrate the cancellation of the $\varepsilon$ poles
for the vector form factor $F^V_1$ in Table~1. 
\begin{table}[h]
\label{tab:pole_cancellation}
  \centerline{  
  \begin{tabular}{ccc}
  \hline\hline 
  $O(\varepsilon^{-2})$ &
  $O(\varepsilon^{-1})$ &
  $O(1)$ \\
  \hline
  $(0.08 \pm 1.2)\cdot 10^{-4}$  &   $(0.68 \pm 11.9)\cdot 10^{-4}$ &  $286.17\pm 0.12$\\
  \hline\hline
  \end{tabular}
  }  
\caption{
  Numerical cancellation of the $1/\varepsilon$ poles in the $O(a^3)$ coefficient
  of the form factor $F^V_1$ for $N_c=3$, $n_f=0$, and $\bar{\mu}=\mu$.
  }
\end{table}  
We observe from Table~1 that the coefficients of the $1/\varepsilon^j$ poles are suppressed relative to the $O(1)$ term by 4 to 5 orders
of magnitude. On the other hand, the absolute error grows from $10^{-4}$ for the $O(\varepsilon^{-2})$ term to
$10^{-1}$ for the $O(1)$ one.
We compare our results at the one-loop level with Refs.~\cite{Sturm:2009kb,Gracey:2010ci}
and at the two-loop level with Ref.~\cite{Gracey:2011fb}. We find agreement with these papers,
however, with one caveat: if we understand the form factors $\Sigma^{S,V,T}_{(j)}$
in Refs.~\cite{Sturm:2009kb,Gracey:2011fb} as being evaluated for the matrix element
$\langle \psi(q)\,O(-q-p)\,\bar{\psi}(p) \rangle$ instead of
$\langle \psi(p)\,O(-q-p)\,\bar{\psi}(q) \rangle$, then we find
$F^{S,V,T}_j=-\Sigma^{S,V,T}_{(j)}$ through two loops.

We now provide all the form factors in Landau gauge, renormalized in the
$\overline{\rm MS}$ scheme.
Our results for the general SU($N$) group are listed in the ancillary file,
and those for the SU(3) group, with $C_F=4/3$ and $C_A=3$, are presented here:
\begin{eqnarray}
  F^S_1 &=& 1 + a \big( 0.6455188559544156 \big) \nonumber\\
    &&{}+ a^2 \big( 48.48858821140752 - 6.346872802912313 n_f \big) \nonumber\\
    &&{}+ a^3 \big( 2396.16(7) - 417.872(2) n_f + 8.6453(1) n_f^2 \big) \,,\\
  F^S_2 &=& a \big( 1.0417365505286484 \big) \nonumber\\
    &&{}+ a^2 \big(11.166805203854107 - 0.462994022457177 n_f \big) \nonumber\\ 
    &&{}+ a^3 \big( 340.47(8) - 65.674(2) n_f + 0.87117(2) n_f^2 \big) \,,\\
  F^V_1 &=& 1 + a \big( - 1.6249301161380183 \big) \nonumber\\
    &&{}+ a^2 \big( -6.1248320773321865 - 0.2362586437086281 n_f \big) \nonumber\\
    &&{}+ a^3 \big( 286.17(12) - 47.915(2) n_f + 2.37045(6) n_f^2 \big) \,, \\
  F^V_2 &=& a \big( -1.4720824544982587 \big) \nonumber\\
    &&{}+ a^2 \big( -18.797490826183573 + 1.3299518483852129 n_f \big) \nonumber\\
    &&{}+ a^3 \big(-658.77(12) + 120.165(2) n_f - 2.13077(4) n_f^2 \big) \,, \\
  F^V_3 &=& a \big( -1.7777777777777777 \big) \nonumber\\
    &&{}+ a^2 \big( -44.38058543042832 + 2.8641975308641974 n_f \big) \nonumber\\
    &&{}+ a^3 \big( -1748.85(6) + 276.890(3) n_f - 6.0501(3) n_f^2 \big) \,, \\  
  F^V_4 &=& F^V_3 \,,\\
  F^V_5 &=& F^V_2 \,,\\  
  F^V_6 &=& a \big( 2.083473101057297 \big) \nonumber\\
    &&{}+ a^2 \big( 39.7873695861749 - 3.009461145971651 n_f \big) \nonumber\\
    &&{}+ a^3 \big(1501.31(4) - 269.099(2) n_f +  6.37227(3) n_f^2 \big) \,,\\
  F^T_1 &=& 1 + a \big( - 0.06232529034504568 \big) \nonumber\\
    &&{}+ a^2 \big( -17.00995398724325 + 1.6001145201364706 n_f \big) \nonumber\\
    &&{}+ a^3 \big( -482.80(7) + 57.814(2) n_f - 0.36852(5) n_f^2 \big) \,,\\
  F^T_2 &=& a \big( -3.125209651585945 \big) \nonumber\\
    &&{}+ a^2 \big( -76.20220922856834 + 5.555928269486126 n_f \big) \nonumber\\
    &&{}+ a^3 \big( -3002.31(16) + 488.628(7) n_f - 11.87338(5) n_f^2 \big) \,,\\
  F^T_3 &=& 2F^T_4 \,,\\
  F^T_4 &=& a \big( 0.15284766163975952 \big) \nonumber\\
    &&{}+ a^2 \big( 0.9097900360378999 - 0.16983073515528835 n_f \big) \nonumber\\
    &&{}+ a^3 \big( -16.54(9) - 6.074(2) n_f + 0.23416(3) n_f^2 \big) \,,\\
  F^T_5 &=& F^T_4 \,,\\
  F^T_6 &=& 2F^T_4 \,,\\
  F^T_7 &=& 4F^T_4 \,,\\
  F^T_8 &=& -F^T_2 \,,    
\end{eqnarray}
where $a=\alpha_s/(4\pi)$ and the {}'t~Hooft mass $\bar{\mu}$ is taken to be equal
to the subtraction point $\mu$ in the SMOM scheme. We retain the error intervals only for
the three-loop contributions, while the one- and two-loop coefficients can be obtained from
Ref.~\cite{Gracey:2011fb} in analytic form.

From the above results, we also obtain the conversion factors between the $\overline{\rm MS}$ and RI/SMOM schemes
for the quark field and mass, $C_q$ and $C_m$, namely,
\begin{equation}
  \psi_R^{\overline{\rm MS}} = C_q^{\rm RI/SMOM} \, \psi_R^{\rm RI/SMOM}\,, \qquad
  m_R^{\overline{\rm MS}} = C_m^{\rm RI/SMOM} \, m_R^{\rm RI/SMOM}\,.
\end{equation}
The result for $C_q^{\rm RI/SMOM}$ is known, since it was shown in Ref.~\cite{Sturm:2009kb} that $C_q^{\rm RI/SMOM}=C_q^{\rm RI'/MOM}$ as a result of a Ward--Takahashi identity.
The result for $C_m^{\rm RI/SMOM}$ at the three-loop order is new and reads
\begin{eqnarray}
  C_q^{\rm RI/SMOM} &=&1 + a^2 \big( -25.46420605097376 + 2.333333333333333 n_f \big) \nonumber\\
    &&{}+ a^3 \big( -1489.980500 + 246.442650 n_f - 6.460905 n_f^2 \big) \,,\\
  C_m^{\rm RI/SMOM} &=& 1 + a \big( - 0.6455188559544155533 \big) \nonumber\\
    &&{}+ a^2 \big( -22.607687567041 + 4.0135394695790 n_f \big) \nonumber\\
    &&{}+ a^3 \big( -860.28(7) + 164.742(2) n_f - 2.18440(10) n_f^2 \big)\,.
\label{eq:cm}
\end{eqnarray}  

If we use, for an estimation, $\alpha_s/\pi\sim0.1$, which approximately corresponds to $\mu=2$~GeV,
then we obtain for the mass conversion factor, with $n_f=3$,
\begin{equation}
  C_m^{\rm RI/SMOM} = 1 - 0.0161 - 0.0066 - 0.0060 + \dots \,.
\end{equation}
Thus, the three-loop contribution is of the same sign and size as the two-loop one.
  
\section{Conclusion}

In this paper, we have established a framework for the evaluation of the
bilinear quark operators of QCD in the MOM scheme at the symmetric kinematic
point.
The results have been obtained for the scalar, vector, and tensor currents with
general tensor and Dirac structures in the three-loop approximation.
The conversion factors between the $\overline{\rm MS}$ and RI/SMOM schemes have been given to three loops as well. The three-loop corrections appear to be comparable in size with the two-loop ones.

\section*{Note added in proof}

After submission, a preprint \cite{Bednyakov:2020ugu} has appeared in which our Eq.~(\ref{eq:cm}) is confirmed and presented in analytic form.

\section*{Acknowledgments}

This work was supported in part by DFG Research Unit FOR~2926 through Grant
No.\ 409651613.
O.L.V. is grateful to V.~Braun and M.~G\"ockeler for fruitfull discussions
and to the University of Hamburg for the warm hospitality.


\bibliography{mybibfile}

\begin{thebibliography}{10}
\expandafter\ifx\csname url\endcsname\relax
  \def\url#1{\texttt{#1}}\fi
\expandafter\ifx\csname urlprefix\endcsname\relax\def\urlprefix{URL }\fi
\expandafter\ifx\csname href\endcsname\relax
  \def\href#1#2{#2} \def\path#1{#1}\fi

\bibitem{Martinelli:1994ty}
G.~Martinelli, C.~Pittori, C.~T. Sachrajda, M.~Testa, A.~Vladikas, {A general
  method for non-perturbative renormalization of lattice operators}, Nucl.
  Phys. B445 (1995) 81--108.
\newblock \href {http://arxiv.org/abs/hep-lat/9411010}
  {\path{arXiv:hep-lat/9411010}}, \href
  {http://dx.doi.org/10.1016/0550-3213(95)00126-D}
  {\path{doi:10.1016/0550-3213(95)00126-D}}.

\bibitem{Sturm:2009kb}
C.~Sturm, Y.~Aoki, N.~H. Christ, T.~Izubuchi, C.~T.~C. Sachrajda, A.~Soni,
  {Renormalization of quark bilinear operators in a momentum-subtraction scheme
  with a nonexceptional subtraction point}, Phys. Rev. D80 (2009) 014501.
\newblock \href {http://arxiv.org/abs/0901.2599} {\path{arXiv:0901.2599}},
  \href {http://dx.doi.org/10.1103/PhysRevD.80.014501}
  {\path{doi:10.1103/PhysRevD.80.014501}}.

\bibitem{Chetyrkin:1999pq}
K.~G. Chetyrkin, A.~R\'etey, {Renormalization and running of quark mass and
  field in the regularization invariant and $\overline{\mathrm{MS}}$ schemes at
  three and four loops}, Nucl. Phys. B583 (2000) 3--34.
\newblock \href {http://arxiv.org/abs/hep-ph/9910332}
  {\path{arXiv:hep-ph/9910332}}, \href
  {http://dx.doi.org/10.1016/S0550-3213(00)00331-X}
  {\path{doi:10.1016/S0550-3213(00)00331-X}}.

\bibitem{Gracey:2003yr}
J.~A. Gracey, {Three loop anomalous dimension of non-singlet quark currents in
  the RI' scheme}, Nucl. Phys. B662 (2003) 247--278.
\newblock \href {http://arxiv.org/abs/hep-ph/0304113}
  {\path{arXiv:hep-ph/0304113}}, \href
  {http://dx.doi.org/10.1016/S0550-3213(03)00335-3}
  {\path{doi:10.1016/S0550-3213(03)00335-3}}.

\bibitem{Gracey:2010ci}
J.~A. Gracey, {RI'/SMOM scheme amplitudes for deep inelastic scattering
  operators at one loop in QCD}, Phys. Rev. D83 (2011) 054024.
\newblock \href {http://arxiv.org/abs/1009.3895} {\path{arXiv:1009.3895}},
  \href {http://dx.doi.org/10.1103/PhysRevD.83.054024}
  {\path{doi:10.1103/PhysRevD.83.054024}}.

\bibitem{Almeida:2010ns}
L.~G. Almeida, C.~Sturm, {Two-loop matching factors for light quark masses and
  three-loop mass anomalous dimensions in the regularization invariant
  symmetric momentum-subtraction schemes}, Phys. Rev. D82 (2010) 054017.
\newblock \href {http://arxiv.org/abs/1004.4613} {\path{arXiv:1004.4613}},
  \href {http://dx.doi.org/10.1103/PhysRevD.82.054017}
  {\path{doi:10.1103/PhysRevD.82.054017}}.

\bibitem{Gorbahn:2010bf}
M.~Gorbahn, S.~J{\"a}ger, {Precise $\overline{\mathrm{MS}}$ light-quark masses
  from lattice QCD in the regularization invariant symmetric
  momentum-subtraction scheme}, Phys. Rev. D82 (2010) 114001.
\newblock \href {http://arxiv.org/abs/1004.3997} {\path{arXiv:1004.3997}},
  \href {http://dx.doi.org/10.1103/PhysRevD.82.114001}
  {\path{doi:10.1103/PhysRevD.82.114001}}.

\bibitem{Gracey:2011fb}
J.~A. Gracey, {RI'/SMOM scheme amplitudes for quark currents at two loops},
  Eur. Phys. J. C71 (2011) 1567.
\newblock \href {http://arxiv.org/abs/1101.5266} {\path{arXiv:1101.5266}},
  \href {http://dx.doi.org/10.1140/epjc/s10052-011-1567-8}
  {\path{doi:10.1140/epjc/s10052-011-1567-8}}.

\bibitem{Gracey:2011zn}
J.~A. Gracey, {Two loop renormalization of the $n = 2$ Wilson operator in the
  RI'/SMOM scheme}, JHEP 03 (2011) 109.
\newblock \href {http://arxiv.org/abs/1103.2055} {\path{arXiv:1103.2055}},
  \href {http://dx.doi.org/10.1007/JHEP03(2011)109}
  {\path{doi:10.1007/JHEP03(2011)109}}.

\bibitem{Gracey:2011zg}
J.~A. Gracey, {Amplitudes for the $n = 3$ moment of the Wilson operator at two
  loops in the RI/'SMOM scheme}, Phys. Rev. D84 (2011) 016002.
\newblock \href {http://arxiv.org/abs/1105.2138} {\path{arXiv:1105.2138}},
  \href {http://dx.doi.org/10.1103/PhysRevD.84.016002}
  {\path{doi:10.1103/PhysRevD.84.016002}}.

\bibitem{Gracey:2020rok}
J.~A. Gracey, {Symmetric point flavour singlet axial vector current
  renormalization at two loops}\href {http://arxiv.org/abs/2001.11282}
  {\path{arXiv:2001.11282}}.

\bibitem{Bednyakov:2020cdf}
A.~Bednyakov, A.~Pikelner, {Four-loop QCD MOM beta functions from the
  three-loop vertices at the symmetric point}\href
  {http://arxiv.org/abs/2002.02875} {\path{arXiv:2002.02875}}.

\bibitem{Chetyrkin:1981qh}
K.~G. Chetyrkin, F.~V. Tkachov, {Integration by parts: The algorithm to
  calculate $\beta$-functions in 4 loops}, Nucl. Phys. B192 (1981) 159--204.
\newblock \href {http://dx.doi.org/10.1016/0550-3213(81)90199-1}
  {\path{doi:10.1016/0550-3213(81)90199-1}}.

\bibitem{Smirnov:2014hma}
A.~V. Smirnov, {FIRE5: A C++ implementation of Feynman Integral REduction},
  Comput. Phys. Commun. 189 (2015) 182--191.
\newblock \href {http://arxiv.org/abs/1408.2372} {\path{arXiv:1408.2372}},
  \href {http://dx.doi.org/10.1016/j.cpc.2014.11.024}
  {\path{doi:10.1016/j.cpc.2014.11.024}}.

\bibitem{Chetyrkin:2006dh}
K.~G. Chetyrkin, M.~Faisst, C.~Sturm, M.~Tentyukov, {$\epsilon$-finite basis of
  master integrals for the integration-by-parts method}, Nucl. Phys. B742
  (2006) 208--229.
\newblock \href {http://arxiv.org/abs/hep-ph/0601165}
  {\path{arXiv:hep-ph/0601165}}, \href
  {http://dx.doi.org/10.1016/j.nuclphysb.2006.02.030}
  {\path{doi:10.1016/j.nuclphysb.2006.02.030}}.

\bibitem{Faisst:2006sr}
M.~Faisst, P.~Maierh{\"o}fer, C.~Sturm, {Standard and $\epsilon$-finite master
  integrals for the $\rho$-parameter}, Nucl. Phys. B766 (2007) 246--268.
\newblock \href {http://arxiv.org/abs/hep-ph/0611244}
  {\path{arXiv:hep-ph/0611244}}, \href
  {http://dx.doi.org/10.1016/j.nuclphysb.2006.12.014}
  {\path{doi:10.1016/j.nuclphysb.2006.12.014}}.

\bibitem{Usyukina:1994iw}
N.~I. Ussyukina, A.~I. Davydychev, {New results for two-loop off-shell
  three-point diagrams}, Phys. Lett. B332 (1994) 159--167.
\newblock \href {http://arxiv.org/abs/hep-ph/9402223}
  {\path{arXiv:hep-ph/9402223}}, \href
  {http://dx.doi.org/10.1016/0370-2693(94)90874-5}
  {\path{doi:10.1016/0370-2693(94)90874-5}}.

\bibitem{Birthwright:2004kk}
T.~G. Birthwright, E.~W.~N. Glover, P.~Marquard, {Master integrals for massless
  two-loop vertex diagrams with three offshell legs}, JHEP 09 (2004) 042.
\newblock \href {http://arxiv.org/abs/hep-ph/0407343}
  {\path{arXiv:hep-ph/0407343}}, \href
  {http://dx.doi.org/10.1088/1126-6708/2004/09/042}
  {\path{doi:10.1088/1126-6708/2004/09/042}}.

\bibitem{Remiddi:1999ew}
E.~Remiddi, J.~A.~M. Vermaseren, {Harmonic polylogarithms}, Int. J. Mod. Phys.
  A15 (2000) 725--754.
\newblock \href {http://arxiv.org/abs/hep-ph/9905237}
  {\path{arXiv:hep-ph/9905237}}, \href
  {http://dx.doi.org/10.1142/S0217751X00000367}
  {\path{doi:10.1142/S0217751X00000367}}.

\bibitem{Kniehl:2017ikj}
B.~A. Kniehl, A.~F. Pikelner, O.~L. Veretin, {Three-loop massive tadpoles and
  polylogarithms through weight six}, JHEP 08 (2017) 024.
\newblock \href {http://arxiv.org/abs/1705.05136} {\path{arXiv:1705.05136}},
  \href {http://dx.doi.org/10.1007/JHEP08(2017)024}
  {\path{doi:10.1007/JHEP08(2017)024}}.

\bibitem{Binoth:2000ps}
T.~Binoth, G.~Heinrich, {An automatized algorithm to compute infrared divergent
  multi-loop integrals}, Nucl. Phys. B585 (2000) 741--759.
\newblock \href {http://arxiv.org/abs/hep-ph/0004013}
  {\path{arXiv:hep-ph/0004013}}, \href
  {http://dx.doi.org/10.1016/S0550-3213(00)00429-6}
  {\path{doi:10.1016/S0550-3213(00)00429-6}}.

\bibitem{Binoth:2003ak}
T.~Binoth, G.~Heinrich, {Numerical evaluation of multi-loop integrals by sector
  decomposition}, Nucl. Phys. B680 (2004) 375--388.
\newblock \href {http://arxiv.org/abs/hep-ph/0305234}
  {\path{arXiv:hep-ph/0305234}}, \href
  {http://dx.doi.org/10.1016/j.nuclphysb.2003.12.023}
  {\path{doi:10.1016/j.nuclphysb.2003.12.023}}.

\bibitem{Smirnov:2015mct}
A.~V. Smirnov, {FIESTA4: Optimized Feynman integral calculations with GPU
  support}, Comput. Phys. Commun. 204 (2016) 189--199.
\newblock \href {http://arxiv.org/abs/1511.03614} {\path{arXiv:1511.03614}},
  \href {http://dx.doi.org/10.1016/j.cpc.2016.03.013}
  {\path{doi:10.1016/j.cpc.2016.03.013}}.

\bibitem{Hahn:2004fe}
T.~Hahn, {{\sc Cuba}---a library for multidimensional numerical integration},
  Comput. Phys. Commun. 168 (2005) 78--95.
\newblock \href {http://arxiv.org/abs/hep-ph/0404043}
  {\path{arXiv:hep-ph/0404043}}, \href
  {http://dx.doi.org/10.1016/j.cpc.2005.01.010}
  {\path{doi:10.1016/j.cpc.2005.01.010}}.

\bibitem{Bednyakov:2020ugu}
A.~Bednyakov, A.~Pikelner, {Quark masses: NNLO bridge from ${\rm RI/SMOM}$ to
  ${\rm \overline{MS}}$ scheme}\href {http://arxiv.org/abs/2002.12758}
  {\path{arXiv:2002.12758}}.

\end{thebibliography}

\end{document}